\documentclass{llncs}
\usepackage{fontenc}
\usepackage[width=122mm,left=12mm,paperwidth=146mm,height=193mm,top=12mm,paperheight=217mm]{geometry}
\usepackage{xcolor}
\usepackage{amssymb}
\usepackage{fontawesome5}
\usepackage{amsfonts}
\usepackage{graphicx}
\usepackage{diagbox}
\usepackage[T1]{fontenc}
\usepackage{color, colortbl}
\definecolor{teagreen}{rgb}{0.9, 0.97, 0.90}
\definecolor{wildblueyonder}{rgb}{0.64, 0.68, 0.82}
\usepackage[hyphens]{url}
\usepackage{caption}
\usepackage{subcaption}
\usepackage{color,soul}
\usepackage{amsmath}
\DeclareMathSymbol{\Beta}{\mathalpha}{operators}{"42}
\usepackage[hidelinks]{hyperref}
\usepackage{xcolor,pifont}
\newcommand*\colourcheck[1]{%
  \expandafter\newcommand\csname #1check\endcsname{\textcolor{#1}{\ding{52}}}%
}
\colourcheck{blue}
\colourcheck{green}
\colourcheck{red}
\usepackage{enumitem}
\usepackage{xspace}
\usepackage{multirow}
\usepackage{colortbl}
\usepackage{array}
\usepackage[misc]{ifsym}
\usepackage{booktabs}
\newcolumntype{C}[1]{>{\centering\arraybackslash}p{#1}}
\usepackage{colortbl}
\usepackage{soul}
\usepackage{tabularx,ragged2e}
\newcolumntype{Y}{>{\centering\arraybackslash}X}

\usepackage{graphicx}
\usepackage{xcolor}

\begin{document}

\newcommand{\smallsection}[1]{\noindent {\bf{\smash{#1}}}}
\newtheorem{obs}{\textbf{Observation}}
\newtheorem{prp}{\textbf{Property}}
\newtheorem{dfn}{\textbf{Definition}}
\newtheorem{trm}{\textbf{Theorem}}
\newcommand{\std}{\scriptsize}
\newcommand\red[1]{\textcolor{red}{#1}}
\newcommand\blue[1]{\textcolor{blue}{#1}}
\newcommand\orange[1]{\textcolor{orange}{#1}}
\newcommand\brown[1]{\textcolor{brown}{#1}}
\newcommand\olive[1]{\textcolor{olive}{#1}}
\newcommand\sunwoo[1]{\textcolor{sunwooblue}{#1}}

\newcommand\mybox[1]{
    \rotatebox{90}{
        \parbox[c][3mm][c]{19mm}{\centering \bf \footnotesize \raggedright #1}
    }
}
\newcommand\mV[0]{\textcolor{green}{\faCheck}}
\newcommand\mX[0]{\textcolor{red}{\faTimes}}

\definecolor{lightgray}{gray}{0.97}

\title{Multi-Behavior Recommender Systems: A Survey}
%
%
\author{Kyungho Kim\inst{1} \and
Sunwoo Kim\inst{1} \and
Geon Lee\inst{1} \and Jinhong Jung\inst{2} \and Kijung Shin\inst{1}\textsuperscript{(\Letter)}}
%


\authorrunning{Kim et al.}
%
\institute{Kim Jaechul Graduate School of AI, KAIST, Seoul, Republic of Korea \\
\email{\{kkyungho, kswoo97, geonlee0325, kijungs\}@kaist.ac.kr} 
\and School of Software, Soongsil University, Seoul, Republic of Korea  \\ \email{jinhong@ssu.ac.kr}}
\maketitle              
\begin{abstract}
Traditional recommender systems primarily rely on a single type of user-item interaction, such as item purchases or ratings, to predict user preferences. However, in real-world scenarios, users engage in a variety of behaviors, such as clicking on items or adding them to carts, offering richer insights into their interests.
Multi-behavior recommender systems leverage these diverse interactions to enhance recommendation quality,  and research on this topic has grown rapidly in recent years.
This survey provides a timely review of multi-behavior recommender systems, focusing on three key steps:
(1) \textbf{Data Modeling:} representing multi-behaviors at the input level, (2) \textbf{Encoding:} transforming these inputs into vector representations (i.e., embeddings), and (3)
\textbf{Training:} optimizing machine-learning models. 
We systematically categorize existing multi-behavior recommender systems based on the commonalities and differences in their approaches across the above steps. 
Additionally, we discuss promising future directions for advancing multi-behavior recommender systems.

\end{abstract}
\section{Introduction}
\label{sec:intro}
With advancements in web applications, such as e-commerce and streaming platforms, users can conveniently purchase a wide variety of items online.
Amid the vast array of available items, users often struggle to find those that best meet their needs. 
Recommender systems address this challenge by automatically identifying suitable items, including those users may not have considered.
In such web applications, users often engage through diverse behaviors beyond simply purchasing or rating items.
Actions, such as clicking on items, adding items to shopping carts, and saving items to wish lists, all provide valuable information about user preferences. Traditional recommender systems that focus solely on a single type of interaction may miss these nuanced insights. 

Multi-behavior recommender systems seek to harness this spectrum of user interactions to improve the accuracy and relevance of recommendations. By integrating diverse behavioral data, these systems can capture nuanced user intents and provide more comprehensive personalization. Interest in multi-behavior recommendation has surged in recent years\footnote{The number of peer-reviewed publications on this topic increased from 37 in 2019-2020 to 79 in 2021-2022, and reached 164 in 2023-2024. 
These counts include research articles with "multi-behavior recommendation" or "multi-relation(al) recommendation" in their titles, abstracts, and/or keywords.}, reflecting the complex decision-making processes of users and the need for more sophisticated methods to interpret them. 




In this survey, we provide an extensive review of multi-behavior recommendation techniques, covering a range of modeling strategies and frameworks. We emphasize the critical role of incorporating multiple user behaviors to improve recommendation accuracy and examine the methodologies used to facilitate this integration. 
Specifically, we examine three key steps of multi-behavior recommendation: data modeling, encoding, and training. For each step, we present a systematic categorization of existing approaches, focusing on how they leverage multiple behaviors.

\paragraph{Step 1: Data Modeling.}
Effective data modeling is fundamental for capturing the relationships between multiple user behaviors in recommender systems.
    \begin{itemize}
        \item \textbf{View-Specific Graphs:}  This approach models each behavior type separately, capturing behavior-specific characteristics and interactions.
        \item \textbf{View-Unified Graph:} By integrating multiple behavior types into a single graph, this approach comprehensively represents user-item interactions.
        \item \textbf{View-Unified Sequences:} By incorporating the temporal order of user behaviors, this approach captures the dynamics of user interactions over time, reflecting evolving preferences.
    \end{itemize}

\paragraph{Step 2: Encoding.} Encoding frameworks transform modeled data into user and item representations, which serve as the basis for providing recommendations.
\begin{itemize}
        \item \textbf{Parallel Encoding:} This approach processes each behavior separately or simultaneously, capturing behavior-specific features in parallel.  
        \item \textbf{Sequential Encoding:} 
This approach captures dynamic dependencies across interactions, integrating the progression of user behaviors. 
    \end{itemize}

\paragraph{Step 3: Training.} Training, the process of optimizing multi-behavior recommender systems, can be guided by various training objectives.
\begin{itemize}
        \item \textbf{Main Training Objectives}: These directly target improving recommendation accuracy, and they are often based on sampling strategies. 
        \item \textbf{Auxiliary Training Objectives}: These improve the quality of latent features through auxiliary tasks and/or self-supervised learning techniques. 
    \end{itemize}

\subsubsection{Comparison with Existing Surveys.}
While multi-behavior information has been leveraged by recommender systems across various settings and domains, the existing survey~\cite{chen2023survey} focuses on  {\textit{multi-behavior sequential recommendation}, particularly methods that account for the temporal order of user interactions over time. 
By contrast, our survey broadens the scope by offering a more comprehensive exploration of diverse techniques across multiple contexts, with a focus on data modeling, encoding frameworks, and training objectives.



\section{Preliminaries for Multi-behavior Recommendation}
\label{sec:prem}
\subsection{Task Description}
\subsubsection{Overview.}
The multi-behavior recommendation aims to predict a target behavior (e.g., purchases, likes) by leveraging interactions from both the target behavior itself and auxiliary behaviors (e.g., clicks, add-to-cart, collections, neutral actions, dislikes) (see Figure~\ref{fig:task}).
Since auxiliary behaviors provide valuable signals about potential target behaviors, effectively integrating them is essential for improving recommendation quality.
\footnote{Compared to a method that relies solely on target behavior (LightGCN~\cite{he2020lightgcn}), state-of-the-art multi-behavior recommender systems (spec., MULE~\cite{lee2024mule}) achieve up to a $463\%$ performance gain in terms of the NDCG@10 metric.}

\begin{figure}
\includegraphics[width=\textwidth]{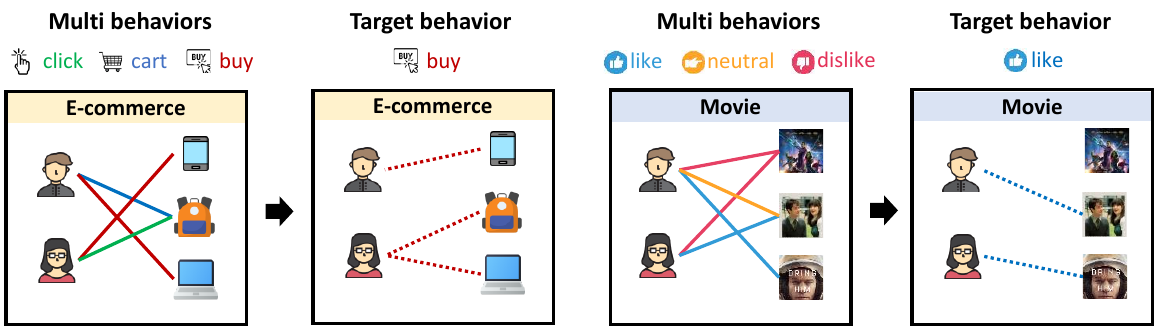}
\caption{Multi-behavior recommendation in two example domains.} \label{fig:task}
\end{figure}
\subsubsection{Key Steps in Multi-behavior Recommendation.} 
We outline the key components of multi-behavior recommendation systems in three main steps.
\begin{enumerate}[start=1,label={\bfseries S\arabic*.}]
    \item \textbf{Data Modeling:} In this step, multi-behavior interactions are represented using specific data structures, such as graphs or sequences, to capture the distinctive characteristics of multi-behaviors or synergy of them (Section~\ref{sec:modeling}).
    \item \textbf{Encoding:} Next, the modeled multi-behaviors are transformed into vector representations (i.e., embeddings) to effectively capture meaningful behavioral patterns (Section~\ref{sec:encoding}).
    \item \textbf{Training:} Finally, the parameters of multi-behavior recommender systems are optimized to enhance their ability to learn and utilize multi-behavior information effectively (Section~\ref{sec:objective}).
\end{enumerate}

We provide an in-depth analysis of current methodologies employed in each of the three phases of multi-behavior recommendation, specifically focusing on how existing methods leverage multi-behavior information. 
Figure~\ref{fig:framework} depicts the comprehensive workflow of multi-behavior recommendation, and Table~\ref{tab:summary} summarizes the key features of existing research in this domain.

\begin{figure}[t]
\includegraphics[width=\textwidth]{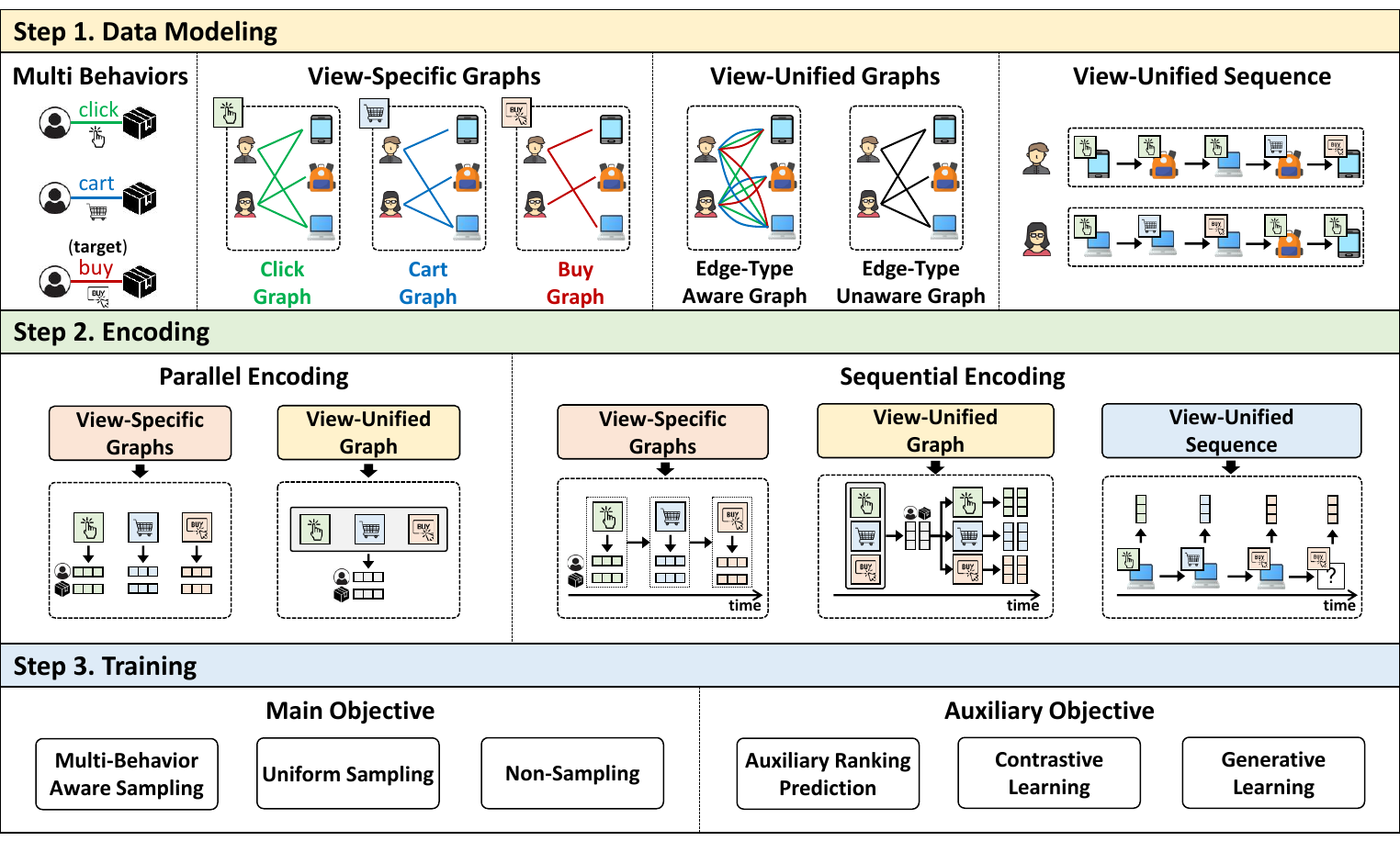}
\caption{Overview of key components in multi-behavior recommendation.} \label{fig:framework}
\end{figure}

\subsection{Benchmark Datasets for Multi-behavior Recommendation}
Datasets are essential for evaluating multi-behavior recommender systems, enabling researchers to analyze their characteristics and identify effective methods for specific domains.
In Table \ref{tab:datasets}, we summarize popular benchmark datasets for multi-behavior recommendation.\footnote{Due to inconsistencies in dataset statistics caused by variations in preprocessing steps, we include the largest reported number of interactions of each dataset.} 

\subsubsection{Dataset Categorization.}
The benchmark datasets are categorized into two groups: \textbf{without explicit feedback} and \textbf{with explicit feedback}. Explicit feedback, such as likes or dislikes~\cite{koren2008factorization}, directly indicates user preferences.

The first category, without explicit feedback, encompasses datasets based solely on implicit feedback, which does not directly reveal user preferences. Examples include datasets containing clicks, cart additions, and purchase histories, commonly found in e-commerce platforms. 

The second category, with explicit feedback, includes datasets where user preferences are directly indicated, such as like or dislike histories. 
This category is common in streaming service platforms.
\subsubsection{Dataset Summaries.} We present a summary of datasets widely used in multi-behavior recommendation, focusing on interaction types and target behaviors.

\paragraph{Datasets without explicit feedback:}
\begin{itemize}
    \item \textbf{Tmall}\footnote{\url{https://tianchi.aliyun.com/dataset/140281}} is a general e-commerce dataset from Alibaba, containing four types of user behaviors: click, collect, cart, and purchase. 
    \item \textbf{Taobao}\footnote{\url{https://tianchi.aliyun.com/dataset/649}} is a general e-commerce dataset from Alibaba, containing three types of user behaviors: click, cart, and purchase.
    \item \textbf{Beibei}\footnote{\url{https://www.beibei.net/}} is a baby-supply e-commerce dataset from BeiBei, containing three types of user behaviors: click, cart, and purchase.
    \item \textbf{JData}\footnote{\url{https://global.jd.com/}} is a general e-commerce dataset from JD, containing four types of user behaviors: click, collect, cart, and purchase. 
    \item \textbf{Online Reatail}\footnote{\url{https://github.com/akaxlh/KHGT}} is a general e-commerce dataset with four types of user behaviors: click, collect (add-to-favorite), cart, and purchase. 
    \item \textbf{IJCAI}\footnote{\url{https://ijcai-15.org/repeat-buyers-prediction-competition/}}
    is a general e-commerce dataset used for an IJCAI competition. It has four types of behaviors: click, add-to-favorite, add-to-cart, and purchase. 
\end{itemize}
In all the above datasets, `purchase' is typically considered the target behavior.
\paragraph{Datasets with explicit feedback:}
\begin{itemize}
    \item \textbf{Yelp}\footnote{\url{https://www.kaggle.com/datasets/yelp-dataset/yelp-dataset}} is a business-review dataset containing four types of user behaviors: dislike, neutral, like, and tip (i.e., user-provided feedback). Ratings ($r$) are categorized into three groups: dislike ($r \leq 2$), neutral ($2 < r < 4$), and like ($r \geq 4$), with like serving as the target behavior. 
    \item \textbf{ML10M}\footnote{\url{https://grouplens.org/datasets/movielens/10m/}} is a movie rating dataset from MovieLens, categorizing user behaviors based on ratings into three groups: dislike ($r \leq 2$), neutral ($2 < r < 4$), and like ($r \geq 4$), with like serving as the target. 
\end{itemize}

\begin{table}
\centering
\caption{
    Comparison of multi-behavior recommender systems. 
}\label{tab:summary}
\scalebox{0.8}{%
\rowcolors{2}{white}{lightgray}
\renewcommand{\arraystretch}{1.2}
\begin{tabularx}{135mm}{l|YYYY|YYY|YYYY}
\hline
\toprule
 \rowcolor{white} & \multicolumn{4}{c|}{\bf Modeling}  
 & \multicolumn{3}{c|}{\bf Encoding} & \multicolumn{3}{c}{\bf Objective}\\
\cmidrule(lr){2-5} \cmidrule(l){6-8} \cmidrule(l){9-11}
\rowcolor{white} \textbf{Model}  & 
\mybox{View-Spec. Graphs}& 
\mybox{View-Uni. Graph} & 
\mybox{View-Uni. Sequences} & 
\mybox{Others} & 
\mybox{Parallel} & 
\mybox{Sequential} & 
\mybox{Others} & 
\mybox{Aux. Rank.} & 
\mybox{Contrastive} & 
\mybox{Generative}\\
\midrule
CMF~\cite{singh2008relational} & \mV & \mX & \mX & \mX & \mV & \mX & \mX & \mX & \mX & \mX \\
MR-BPR~\cite{Krohn-Grimberghe2012multi} & \mV & \mX & \mX & \mX & \mV & \mX & \mX & \mX & \mX & \mX \\
MC-BPR~\cite{loni2016bayesian} & \mV & \mX & \mX & \mX & \mV & \mX & \mX & \mX & \mX & \mX \\
BPRH~\cite{qiu2018bprh} & \mX & \mX & \mX & \mV & \mX & \mX & \mV & \mX & \mX & \mX \\
VALS~\cite{ding2018improving} & \mV & \mX & \mX & \mX & \mV & \mX & \mX & \mX & \mX & \mX\\
BINN~\cite{li2018learning} & \mX & \mX & \mV & \mX & \mX & \mV & \mX & \mX & \mX & \mX \\
 NMTR~\cite{gao2019neural} & \mV & \mX & \mX & \mX & \mX & \mV & \mX & \mV & \mX & \mX \\
DIPN~\cite{guo2019buying} & \mX & \mV & \mX & \mX & \mV & \mX & \mX & \mV & \mX & \mX \\
 MRMN~\cite{zhou2019collaborative} &\mV & \mX & \mX & \mX & \mV & \mX & \mX & \mV & \mX & \mX \\
MATN~\cite{xia2020multiplex} & \mX & \mV & \mX & \mX & \mV & \mX & \mX & \mV & \mX & \mX \\
 MBGCN~\cite{jin2020multi} & \mX & \mV & \mX & \mX & \mV & \mX & \mX & \mX & \mX & \mX \\
EHCF~\cite{chen2020efficient} & \mV & \mX & \mX & \mX & \mV & \mX & \mX & \mV & \mX & \mX \\
 MGNN~\cite{zhang2020multiplex} & \mX & \mV & \mX & \mX & \mV & \mX & \mX & \mV & \mX & \mX\\
GMNR~\cite{xia2021multi} & \mX & \mV & \mX & \mX & \mV & \mX & \mX & \mV & \mX & \mX \\
 MB-GMN~\cite{xia2021graph} & \mX & \mV & \mX & \mX & \mV & \mX & \mX & \mV & \mX & \mX \\
HMG-CR~\cite{yang2021hyper} & \mX & \mX & \mX & \mV & \mV & \mX & \mX & \mV & \mV & \mX \\
 GHCF~\cite{chen2021graph} & \mX & \mV & \mX & \mX & \mV & \mX & \mX & \mV & \mX & \mX \\
KHGT~\cite{xia2021knowledge} & \mX & \mV & \mX & \mV & \mV & \mX & \mX & \mV & \mX & \mX \\
 S-MBRec~\cite{gu2022self} & \mV & \mX & \mX & \mX & \mV & \mX & \mX & \mV & \mV & \mX \\
CML~\cite{wei2021contrastive} & \mV & \mX & \mX & \mX & \mV & \mX & \mX & \mV & \mV & \mX \\
 MB-STR~\cite{yuan2022multi} & \mX & \mV & \mX & \mX & \mX & \mV & \mX & \mX & \mX & \mV\\
MBHT~\cite{yang2022multi} & \mX & \mX & \mV & \mV & \mX & \mX & \mX & \mX & \mX & \mV \\
 MMCLR~\cite{wu2022multi} & \mX & \mV & \mV & \mX & \mX & \mV & \mX & \mX & \mV & \mX \\
 NextIP~\cite{luo2022dual} & \mX & \mV & \mX & \mX & \mX & \mV & \mX & \mX & \mV & \mV\\
CKML~\cite{meng2023coarse} & \mX & \mV & \mX & \mV & \mV & \mX & \mX & \mV & \mX & \mX \\
 CRGCN~\cite{meng2023coarse} & \mV & \mX & \mX & \mX & \mX & \mV & \mX & \mV & \mX & \mX \\
MB-CGCN~\cite{cheng2023multi}  & \mV & \mX & \mX & \mX & \mX & \mV & \mX & \mX & \mX & \mX \\
 PKEF~\cite{meng2023parallel} & \mV & \mX & \mX & \mX & \mV & \mX & \mX & \mV & \mX & \mX \\
HPMR~\cite{meng2023hierarchical} & \mV & \mX & \mX & \mX & \mV & \mX & \mX & \mV & \mX & \mX \\
 CHCF~\cite{luo2023criterion} & \mV & \mX & \mX & \mX & \mV & \mX & \mX & \mV & \mX & \mX \\
MB-HGCN~\cite{yan2023mbhgcn} & \mV & \mV & \mX & \mX & \mX & \mV & \mX & \mV & \mX & \mX \\
 PBAT~\cite{su2023personalized} & \mX & \mX & \mV & \mX & \mX & \mV & \mX & \mV & \mX & \mX \\
KMVLR~\cite{xuan2023knowledge} & \mV & \mX & \mX & \mV & \mV & \mX & \mX & \mX & \mV & \mX \\
 MBA~\cite{xin2023improving} & \mX & \mX & \mX &\mV & \mX & \mX & \mV & \mX & \mX & \mX \\ 
MBSSL~\cite{xu2023multi} & \mV & \mX & \mX & \mX & \mV & \mX & \mX & \mV & \mV & \mX \\
 BCIPM~\cite{yan2024behavior} & \mV & \mX & \mX & \mX & \mV & \mX & \mX & \mV & \mX & \mX \\
END4Rec~\cite{han2024efficient} & \mX & \mX & \mV & \mX & \mX & \mV & \mX & \mX & \mV & \mX \\
 MBGen~\cite{liu2024multi} & \mX & \mX & \mV & \mX & \mX & \mV & \mX & \mX & \mX & \mV \\
MULE~\cite{lee2024mule} & \mV & \mV & \mX & \mX & \mX & \mV & \mX & \mX & \mX & \mX \\
 DA-GCN~\cite{zhu2024multi} & \mX & \mX & \mX & \mV & \mV & \mX & \mX & \mV & \mX & \mX \\
HMAR~\cite{elsayed2024hmar} & \mX & \mX & \mV & \mX & \mX & \mV & \mX & \mV & \mX & \mX \\
 PO-GCN~\cite{zhang2024multi} & \mX & \mX & \mX & \mV & \mV & \mX & \mX & \mV & \mX & \mX \\
\bottomrule
\hline
\end{tabularx}
}
\end{table}



\begin{table}
    \centering
    \caption{Summary of datasets used for multi-behavior recommendation. Behaviors providing implicit feedback are highlighted in \textbf{bold}, while those providing explicit feedback are \underline{underlined}.
    Target behaviors are denoted by ${\dagger}$.
    }
    \label{tab:datasets}
    \scalebox{0.93}{%
    \begin{tabular}{l  c c c}
        \toprule
         \textbf{Dataset} & \textbf{Field} & \textbf{Types of Behaviors} & \textbf{\# of Interactions}\\
        \midrule
         Tmall & E-commerce & \ \textbf{Click}, \textbf{Collect}, \textbf{Cart}, \textbf{Purchase}$^{\dagger}$ \ & $2.3 \times 10^6$ \\
         Taobao & E-commerce & \textbf{Click}, \textbf{Cart}, \textbf{Purchase}$^{\dagger}$ & $7.6 \times 10^6$\\
         Beibei & E-commerce & \textbf{Click}, \textbf{Cart}, \textbf{Purchase}$^{\dagger}$ & $3.5 \times 10^6$\\
         JData & E-commerce & \textbf{Click}, \textbf{Collect}, \textbf{Cart}, \textbf{Purchase}$^{\dagger}$ & $2.2 \times 10^6$\\
         Online Retail & E-commerce & \textbf{Click}, \textbf{Collect}, \textbf{Cart}, \textbf{Purchase}$^{\dagger}$ & $6.4 \times 10^7$\\
         IJCAI & E-commerce & \textbf{Click}, \textbf{Collect}, \textbf{Cart}, \textbf{Purchase}$^{\dagger}$ & $3.6 \times 10^7$\\
        \midrule
     Yelp & \ Business Reviews \ & \textbf{Tip}, \underline{Dislike}, \underline{Neural}, \underline{Like}$^{\dagger}$ & $1.4 \times 10^6$\\
         ML10M & Movie Ratings & \underline{Dislike}, \underline{Neutral}, \underline{Like}$^{\dagger}$ & $9.9 \times 10^6$\\
        \bottomrule
    \end{tabular}
    }
\end{table}


\section{Step 1. Data Modeling}
\label{sec:modeling}
In this and the next two sections, we review recommender systems, focusing on three key steps: \textbf{data modeling}, \textbf{encoding}, and \textbf{training}.

The initial step, data modeling, represents multi-behavior interactions as input-level data structures, such as graphs or sequences.
The key challenge of this step lies in how to express rich and diverse interactions effectively. 
Broadly, there are three categories: 
(a) \textbf{view-specific graphs}, which model each behavior with distinct graphs; 
(b) \textbf{view-unified graph}, which models various behaviors with a single unified graph; and
(c) \textbf{view-unified sequences}, which model all behaviors of each user with a single sequence. 
We describe each category below.

\subsection{View-Specific Graphs} 
The view-specific graphs modeling approach represents each type of user interaction as a separate graph.\footnote{We consider single-matrix representation as a type of single-graph representation.}
This method independently models the relationships for each behavior, aiming to preserve the unique characteristics of each interaction type. 
For instance, as illustrated in Figure~\ref{fig:framework}, clicks, add-to-cart actions, and purchases can be expressed as three distinct graphs, each corresponding to a specific behavior type.
View-specific graphs can be further classified 
depending on whether the order between different behaviors is considered.

\subsubsection{Non-Ordered View-Specific Graphs.} 
Many studies emphasize the importance of preserving the unique information of each behavior.
To achieve this, view-specific graphs are processed in parallel, without considering any order between them~\cite{singh2008relational,loni2016bayesian,Krohn-Grimberghe2012multi,ding2018improving,gao2019neural,chen2020efficient,gu2022self,meng2023hierarchical,meng2023parallel,wei2022constrastive,zhou2019collaborative,xin2023improving,zhang2020multiplex,luo2023criterion}. 
This parallel modeling ensures that the unique properties of each behavior are captured effectively.

\subsubsection{Ordered View-Specific Graphs.} Some studies~\cite{gao2019neural,yan2023cascading,cheng2023multi} incorporate an inherent cascading order of user behaviors, such as first clicking an item, then adding it to a cart, and eventually purchasing it.
To capture the progression of user behaviors, the view-specific graphs are assigned an order that aligns with the behavior sequence.
Later, recommender systems encode these graphs based on the assigned order, which will be elaborated on in Section~\ref{subsec:seqencoding}.

\subsection{View-Unified Graph} 
The view-unified graph modeling approach
consolidates diverse user behaviors into a single graph, capturing their interactions and synergies. 
This approach can be further categorized into two sub-approaches based on how they handle different edge types within the graph.

\subsubsection{Edge-Type-Unaware Unified Graph.} The edge-type-unaware unified graph approach aggregates all user behaviors into a single homogeneous graph, treating them as identical types of relationships between users and items~\cite{yan2024behavior,yan2023mbhgcn,lee2024mule}. 
This simple approach offers several computational advantages and enables the direct extension of the well-known homogeneous graph representation technique (e.g., renormalization trick~\cite{kipf2016semi}).

\subsubsection{Edge-Type-Aware Unified Graph.} 
The edge-type-aware unified graph approach preserves the type of each edge by leveraging a heterogeneous graph representation~\cite{jin2020multi,xia2020multiplex,zhang2020multiplex,xia2021graph,xia2021multi,chen2021graph}.
For instance, each edge contains information on whether the corresponding edge expresses purchase or click.
By distinguishing edge types, the graph can effectively capture the unique properties of each behavior while exhibiting the synergy between different interactions at the same time.
Upon this heterogeneous graph, message-passing techniques specifically tailored to handle multiple edge types (e.g., HetGNN~\cite{zhang2019heterogeneous}) are leveraged, which will be elaborated on in Section~\ref{sec:encoding}.

\subsection{View-Unified Sequences} 
The view-unified sequence approach is distinct from the previous graph-based modeling techniques in that it explicitly incorporates the sequential nature of user interactions. Unlike methods that ignore the temporal order of user-item interactions, this modeling aims to capture the dynamics of user behaviors over time. This approach leverages a unified sequence that includes all types of behaviors, considering both the order in which they occurred and the specific type of interaction (i.e., behavior). In this method, the view-unified sequence is used as an input to predict the next item for a target behavior~\cite{li2018learning,guo2019buying,yuan2022multi,liu2024multi,yang2022multi,luo2022dual,su2023personalized,elsayed2024hmar,wu2022multi,han2024efficient}. From a behavior-aware user-item interaction sequence, the model gains a richer understanding of user intentions, considering both the temporal dependencies and the distinct characteristics of each behavior. 

\subsection{Other Representations} 
In addition to these primary categories, several advanced techniques exploit multi-behavior signals through unique (hyper)graph representations. HMG-CR~\cite{yang2021hyper} constructs a hyper meta-graph using hyper meta-paths, which logically combine multiple meta-path schemas to link nodes in heterogeneous graphs. For richer semantic integration, KGHT~\cite{xia2021knowledge}, CKML~\cite{meng2023coarse}, and KMVLR~\cite{xuan2023knowledge} employ knowledge graphs to model item-item relations by leveraging external knowledge with user-item interactions. DA-GCN~\cite{zhu2024multi} introduces personalized directed acyclic behavior graphs, incorporating all observed behavior paths to better model behavior dependencies. MBHT~\cite{yang2022multi} employs item-wise hypergraphs, capturing item semantic dependencies and user-personalized multi-behavior dependencies for comprehensive interaction modeling. PO-GCN~\cite{zhang2024multi} constructs a unified weighted graph with a graded partial order calculated by a rank function. 

\section{Step 2. Encoding}
\label{sec:encoding}
Once multi-behavior data is modeled, the encoding step transforms the structured data into meaningful representations that can be effectively used for recommendation.
In this section, we categorize encoding techniques into (a) \textbf{parallel encoding} and (b) \textbf{sequential encoding}. 
For each category, we provide the underlying intuition, corresponding frameworks, and representative examples.

\subsection{Parallel Encoding}

The intuition behind parallel encoding is to preserve the unique characteristics of each behavior while enabling collaborative learning across behaviors (either implicitly or explicitly).
It assumes no specific order among behaviors and can process them either independently through separate view-specific graphs or collectively by integrating them into a unified graph that distinguishes between interaction types. 
Existing parallel encoding frameworks can be further categorized into (a) encoding for non-ordered view-specific graphs and (b) encoding for edge-type-aware unified graphs.

\subsubsection{Encoding Non-ordered View-Specific Graphs.} 
In this framework, encoding is performed independently for each behavior-specific graph, ensuring that the unique characteristics of each behavior are preserved without interference from others.
Various encoding techniques have been employed to learn from each view-specific graph~\cite{singh2008relational,loni2016bayesian,gu2022self,meng2023hierarchical,meng2023parallel,wei2022constrastive}.
For example, CMF~\cite{singh2008relational} employs matrix factorization, while more recent methods, such as S-MBRec~\cite{gu2022self} and PKEF~\cite{meng2023parallel}, leverage (homogeneous) graph neural networks (e.g., GCN~\cite{kipf2016semi} and LightGCN~\cite{he2020lightgcn}).
These encoding methods facilitate collaborative learning within each behavior while typically leveraging shared initial embeddings across behaviors.
The shared initial embeddings enable implicit information sharing among different interaction types. 

\subsubsection{Encoding an Edge-Type-Aware Unified Graph.} 
In this framework, interactions from multiple behaviors are integrated into a unified graph, while distinguishing which behavior each interaction originates from~\cite{jin2020multi,xia2020multiplex,zhang2020multiplex,xia2021graph,xia2021multi,chen2021graph}.
The key challenge in encoding such an edge-type-aware unified graph lies in effectively capturing the inter-dependencies between interactions from different behaviors.
For example, MATN~\cite{xia2020multiplex} employs a Transformer network with embeddings from multiple behaviors as input. Specifically, by leveraging shared key and memory slots, MATN captures cross-behavior dependencies and refines type-specific contextual embeddings.
MBGCN~\cite{jin2020multi} applies heterogeneous graph neural networks to propagate information in a unified heterogeneous graph, effectively distinguishing behavior types through edge-type-aware propagation.

\subsection{Sequential Encoding}\label{subsec:seqencoding}
Sequential encoding aims to capture temporal dependencies in multi-behavior interactions.
It either assumes an inherent sequence in behavior interactions or leverages explicit temporal information to better understand these dynamics.
Existing methods within this category can be further divided into (a) \textbf{encoding of ordered view-specific graphs}, (b) \textbf{encoding of an edge-type-unaware unified graph}, and (c) \textbf{encoding of a view-unified sequence}.

\subsubsection{Encoding Ordered View-Specific Graphs.} 
This framework assumes a specific order of behaviors (e.g., click $\rightarrow$ cart $\rightarrow$ purchase). 
Each behavior is modeled as a separate graph, with temporal dependencies explicitly captured to reflect this sequence~\cite{gao2019neural,yan2023cascading,cheng2023multi}.
Typically, the output from the previous behavior is used as input to encode the next behavior.
Each view-specific graph can be encoded using (neural) matrix factorization~\cite{gao2019neural} or homogeneous graph neural networks~\cite{yan2023cascading}.
Residual connections across behaviors~\cite{yan2023cascading} or linear transformations between behaviors~\cite{cheng2023multi} have been used to enhance this approach.

\subsubsection{Encoding an Edge-Type-Unaware Unified Graph.} 
This approach leverages an edge-type-unaware unified graph to generate initial embeddings, which are subsequently used as inputs for encoding behavior-specific information ~\cite{yan2024behavior,yan2023mbhgcn,lee2024mule}.
BIPN~\cite{yan2024behavior}, for instance, uses an edge-type-unaware unified graph to learn general representations and propagates them into a preference filtering network implemented as a simple MLP with three filtering layers to infer user preference.
MULE~\cite{lee2024mule} obtains general representations from an edge-type-unaware unified graph and then propagates them into subgraphs that distinguish (a) auxiliary behaviors intersecting with the target behavior from (b) those that do not. After that, it employs an attention mechanism to discern whether each interaction genuinely indicates user interest.


\subsubsection{Encoding View-Unified Sequences.} 
To encode view-unified sequences, it is important to capture temporal dependencies between behaviors of multiple types.
Various sequential models have been employed~\cite{li2018learning,guo2019buying,yuan2022multi,liu2024multi,yang2022multi,su2023personalized,elsayed2024hmar,wu2022multi}.
For example, DIPN~\cite{guo2019buying} uses bidirectional GRUs~\cite{cho2014learning} combined with hierarchical attention to capture intra-sequence and inter-sequence dependencies, while MMCLR~\cite{wu2022multi} employs Bert4Rec~\cite{sun2019bert4rec} to learn behavior-specific representations over time. 
Transformer-based models, such as MB-STR~\cite{yuan2022multi} and MBGen~\cite{liu2024multi}, use self-attention mechanisms to model heterogeneous dependencies and fine-grained sequential patterns, leading to a deeper understanding of user interactions. End4Rec~\cite{han2024efficient} utilizes Fourier-transformation-based modules to capture behavior patterns effectively, incorporating denoising mechanisms to effectively filter out noise from user behavior data.

\subsection{Others}
Many studies develop unique message-passing frameworks and specialized encoding frameworks to capture the intricacies of multi-behavior interactions. EHCF~\cite{chen2020efficient} uses a transfer mechanism that projects predictions of preceding behaviors (e.g., view) to subsequent behaviors (e.g., purchase) through translation in the embedding space, capturing the inter-dependencies between behavior types. HMG-CR~\cite{yang2021hyper} utilizes GNNs to encode hyper meta-graphs that represent user behavior patterns and capture both semantic relations and topological structures. KHGT~\cite{xia2021knowledge} employs a hierarchical graph transformer architecture that utilizes graph-based message passing combined with self-attention mechanisms and incorporates temporal dynamics through time-aware embeddings. DA-GCN~\cite{zhu2024multi} introduces a directed acyclic graph-based message-passing framework that propagates messages across user-specific and item-specific behavior graphs, refining the embeddings based on cross-behavior dependencies.  EIDP~\cite{chen2024explicit} introduces a dual-path Transformer-based framework that distinguishes explicit and implicit multi-behavior relationships. The explicit modeling path captures intra-behavior dynamics within each behavior type, 
while the implicit modeling path focuses on inter-behavior collaborations across behavior types. 

\subsection{Discussions}
Each of these encoding frameworks presents a unique and effective method for encoding user behaviors in multi-behavior recommender systems. Parallel encoding allows for the detailed modeling of behavior-specific patterns by processing each interaction type separately. However, it does not account for the relationships between behaviors, potentially missing the inter-behavior dependencies. Sequential encoding, in contrast, captures the evolving nature of user interactions, making it well-suited for modeling temporal dynamics and providing tailored recommendations. Its drawback lies in the difficulty of integrating a comprehensive view of user preferences across all behaviors. 

\section{Step 3. Training}
\label{sec:objective}
In multi-behavior recommender systems, designing an effective training objective is crucial for ensuring recommendation relevance. We categorize training objectives in existing works into (a) \textbf{main objectives}, which directly aim to optimize recommendation performance, and (b) \textbf{auxiliary objectives}, which indirectly enhance performance by leveraging additional information. 

\subsection{Main Objectives}
The main objective in multi-behavior recommender systems is to accurately predict the target behavior, which is the primary goal of the recommendation task. 
Achieving this requires designing an effective loss function that guides the model to distinguish between positive and negative samples, ensuring that the scores for positive samples are higher than those for negative samples.~\footnote{Positive samples (or items) are those that users have interacted with, whereas negative items are those they have not.} 

Generally, positive samples are straightforward to define, i.e., they are the items with which users have interacted under the target behavior. 
However, identifying suitable negative samples is more challenging, especially since we lack explicit negative feedback (e.g., dislikes) in many multi-behavior recommendation scenarios. 
Therefore, selecting proper negative samples is critical to effectively learning user preferences. 
To select proper negative samples, the following approaches have been studied.


\subsubsection{Multi-Behavior-Aware Sampling.} Early works~\cite{Krohn-Grimberghe2012multi,loni2016bayesian,qiu2018bprh,li2018learning} on multi-behavior recommendation leverage this strategy to select negative samples that are more informative. For instance, MC-BPR~\cite{loni2016bayesian} proposes a negative sampling strategy tailored for multi-behavior feedback, assigning different sampling probabilities for distinct cases. In particular, it assigns different sampling probabilities between items that have been viewed but not purchased and those that have not been viewed at all.  
BPRH~\cite{qiu2018bprh} introduces an adaptive sampler that considers the co-occurrence of multiple behavior types. 

\subsubsection{Uniform Sampling.} 
In this approach, negative samples for each user are uniformly selected from items with which the user has not interacted under the target behavior. 
Many works adopt this uniform sampling due to its simplicity and efficiency~\cite{zhou2019collaborative,xia2020multiplex,jin2020multi,zhang2020multiplex,xia2021graph,xia2021multi,yang2021hyper,xia2021knowledge,gu2022self,wei2022constrastive,yang2022multi,wu2022multi,yan2023cascading,cheng2023multi,meng2023parallel,yan2023mbhgcn,su2023personalized,xuan2023knowledge,lee2024mule,yan2024behavior,zhu2024multi,elsayed2024hmar}. 

\subsubsection{Non-Sampling.} Non-sampling-based loss functions eliminate the need for negative sampling, 
particularly in the context of heterogeneous interactions. This approach considers all items that a user has not interacted with as negative samples. EHCF~\cite{chen2020efficient} proposes a non-sampling-based loss function derived from a weighted regression loss~\cite{hu2008collaborative}. 
This approach captures multiple behaviors by considering all interactions, rather than relying on sampling.
It effectively addresses data sparsity and improves representation quality by utilizing all available information. 
Other works~\cite{chen2021graph,meng2023hierarchical,luo2023criterion,xu2023multi} have also explored this strategy to effectively consider multiple types of relations.

\subsubsection{Discussions.}
Multi-behavior-aware sampling leverages auxiliary behavioral information to select more informative negative samples, enhancing the model's ability to differentiate among users' various preferences and capture complex user preferences. 
Uniform sampling offers simplicity and efficiency but may include less informative negatives or false negatives. Non-sampling-based loss functions eliminate negative sampling procedures, and they can be particularly beneficial in scenarios with data sparsity and heterogeneous interactions.

\subsection{Auxiliary Objectives} 
To further improve model performance and representation quality, especially when data for the target behavior is sparse, many multi-behavior recommender systems incorporate auxiliary objectives. These objectives leverage additional information from auxiliary behaviors to enhance representation learning and improve model robustness. Broadly, auxiliary objectives can be divided into three categories: (a) \textbf{auxiliary ranking prediction}, (b) \textbf{contrastive learning}, and (c) \textbf{generative learning}.

\subsubsection{Auxiliary Ranking Prediction.}
A widely used auxiliary objective is auxiliary ranking prediction, which applies multi-task learning to predict auxiliary behaviors alongside the target behavior~\cite{gao2019neural,guo2019buying,xia2020multiplex,chen2020efficient,xia2021graph,zhang2020multiplex,meng2023parallel,meng2023hierarchical,chen2021graph,yan2023cascading,yan2023mbhgcn,liu2024multi,zhu2024multi,elsayed2024hmar}. Through the joint learning for both target and auxiliary behaviors, the model captures additional signals embedded in user interactions, allowing it to learn shared patterns and underlying correlations. This joint learning approach leads to improved recommendations by 
leveraging richer behavioral relationships.

\subsubsection{Contrastive Learning.}
Contrastive learning,
which is a widely used self-supervised learning technique,
has been applied to  
various recommender systems \cite{lin2022improving,yu2022graph,yu2023xsimgcl,ma2022crosscbr,yang2022knowledge,Wu2022DCRec,wei2021contrastive}.
It enhances representation learning by effectively capturing meaningful similarities and differences. Contrastive learning has also been adopted in multi-behavior recommender systems~\cite{yang2021hyper,gu2022self,wei2022constrastive,wu2022multi,luo2022dual,xu2023multi,han2024efficient}. This approach typically contrasts embeddings derived from auxiliary behaviors with those from the target behavior to uncover meaningful relationships. For example, S-MBRec~\cite{gu2022self} 
contrasts target and auxiliary behaviors to capture both their commonality and differences, effectively addressing data sparsity while reducing redundancy in auxiliary signals. CML~\cite{wei2022constrastive} aligns type-specific behavior embeddings using contrastive learning while capturing user-specific behavior heterogeneity through a meta-contrastive network. MMCLR~\cite{wu2022multi} combines multi-behavior and multi-view contrastive learning to model both coarse-grained commonalities and fine-grained differences, aligning sequence-based and graph-based user representations while capturing behavior-specific priorities. HMG-CR~\cite{yang2021hyper} employs graph contrastive learning to compare hyper meta-graphs, adaptively modeling multi-behavior dependencies in a progressive and structured manner. MBSSL~\cite{xu2023multi} combines inter-behavior and intra-behavior contrastive learning to transfer knowledge from auxiliary behaviors. End4Rec~\cite{han2024efficient} contrasts three types of sequences (i.e., denoised sequences, original sequences, and noise sequences) to decouple noise from meaningful user preferences.

\subsubsection{Generative Learning.}
Generative learning has gained increasing attention in recommendation tasks~\cite{li2023graph,xia2023automated,zhai2024actions}, including multi-behavior recommender systems. 
It enhances the ability of recommender systems to capture user-item interactions by training them to predict masked items or behaviors~\cite{yuan2022multi,yang2022multi,luo2022dual,han2024efficient,liu2024multi}.
For instance, MB-STR~\cite{yuan2022multi} utilizes behavior-aware masked item prediction to reconstruct missing items in the interactions of a user. 
This approach effectively captures multi-behavior dependencies and sequential patterns while mitigating data sparsity and negative transfer.
Similarly, MBGen~\cite{liu2024multi} employs sequence-to-sequence generative learning, which auto-regressively predicts the next behavior and item in a heterogeneous sequence of user interactions. 

\subsubsection{Others.}
MBA~\cite{xin2023improving} is trained to maximize
the likelihood of observed behavioral data while minimizing the KL divergence between auxiliary and target behavior distributions, enabling effective data denoising and knowledge transfer.



\section{Challenges and Future Directions}
\label{sec:challenge}
In this section, we discuss key challenges and future research directions in multi-behavior recommendation.

\subsubsection{Data Sparsity and Imbalance.} 
In many real-world scenarios, certain types of behaviors are abundant, while others remain sparse.
This imbalance may cause recommender systems to prioritize learning abundant behaviors while failing to capture sparse ones.
Advanced self-supervised learning methods that incorporate external knowledge offer a promising direction for addressing these issues.

\subsubsection{Scalability and Efficiency.} 
Real-world user behaviors often occur at scale, making scalability essential for multi-behavior recommender systems to handle them effectively.
Developing a scalable recommender system along with an efficient training strategy, such as sampling-based techniques, would be a promising research direction.

\subsubsection{Temporal Dynamics.} 
User preferences change over time, making it crucial to capture these dynamics for relevant recommendations. 
A promising research direction involves integrating temporal attention mechanisms or dynamic graph-based methods that adaptively update embeddings to reflect time-sensitive changes in user behavior and the evolving relationships between users.

\subsubsection{Interpretability.} Deep learning-based recommender systems, especially those leveraging complex architectures,
often act as black boxes. 
Enhancing interpretability to explain how different behaviors influence recommendations is a crucial research direction for fostering user trust and transparency.
To achieve this, various techniques, such as attention visualization and counterfactual reasoning, can be leveraged.

\subsubsection{Privacy and Ethical Considerations.} 
Detailed user behavioral data raises privacy concerns, making security and ethical compliance essential.
A key research direction involves exploring privacy-preserving methods, such as federated learning and differential privacy, to ensure the secure training of multi-behavior recommender systems while minimizing data exposure risks.


\section{Conclusions}
\label{sec:conclusion}
In this survey, we provide the first comprehensive review of multi-behavior recommender systems. 
We systematically categorize existing multi-behavior recommendation systems based on their approaches across three key steps: data modeling, encoding, and training. Specifically, we examine how multi-behavior data are modeled in various forms and effectively encoded to capture nuanced relationships across multiple behaviors. Additionally, we investigate the role of primary and auxiliary training objectives in leveraging diverse behavioral signals.
Notably, unlike previous surveys, our work covers multi-behavior recommendation across various contexts beyond sequential recommendation.

By fostering a systematic understanding of multi-behavior recommender systems, from early research to recent advancements, this survey aims to aid researchers and practitioners in developing more sophisticated and effective solutions for multi-behavior recommendation.

\subsubsection*{Acknowledgements.}
This work was partly supported by Institute of Information \& Communications Technology Planning \& Evaluation (IITP) grant funded by the Korea government (MSIT) (No. RS-2024-00438638, EntireDB2AI: Foundations and Software for Comprehensive Deep Representation Learning and Prediction on Entire Relational Databases, 50\%)
(No. RS-2019-II190075, Artificial Intelligence Graduate School Program (KAIST), 10\%).
This work was partly supported by the National Research Foundation of Korea (NRF) grant funded by the Korea government (MSIT) (No. RS-2024-00406985, 40\%).

%
%
%
%
\bibliographystyle{splncs04}
\bibliography{ref}

\begin{thebibliography}{10}
\providecommand{\url}[1]{\texttt{#1}}
\providecommand{\urlprefix}{URL }
\providecommand{\doi}[1]{https://doi.org/#1}

\bibitem{chen2021graph}
Chen, C., Ma, W., Zhang, M., Wang, Z., He, X., Wang, C., Liu, Y., Ma, S.: Graph heterogeneous multi-relational recommendation. In: AAAI (2021)

\bibitem{chen2020efficient}
Chen, C., Zhang, M., Ma, W., Zhang, Y., Liu, Y., Ma, S.: Efﬁcient heterogeneous collaborative filtering without negative sampling for recommendation. In: AAAI (2020)

\bibitem{chen2024explicit}
Chen, M., Pan, W., Ming, Z.: Explicit and implicit modeling via dual-path transformer for behavior set-informed sequential recommendation. In: KDD (2024)

\bibitem{chen2023survey}
Chen, X., Li, Z., Pan, W., Ming, Z.: A survey on multi-behavior sequential recommendation. arXiv preprint arXiv:2308.15701  (2023)

\bibitem{cheng2023multi}
Cheng, Z., Han, S., Liu, F., Zhu, L., Gao, Z., Peng, Y.: Multi-behavior recommendation with cascading graph convolution networks. In: WWW (2023)

\bibitem{cho2014learning}
Cho, K., van Merrienboer, B., Gulcehre, C., Bahdanau, D., Bougares, F., Schwenk, H., Bengio, Y.: Learning phrase representations using rnn encoder-decoder for statistical machine translation. In: EMNLP (2014)

\bibitem{ding2018improving}
Ding, J., Yu, G., He, X., Quan, Y., Li, Y., Chua, T.S., Jin, D., Yu, J.: Improving implicit recommender systems with view data. In: IJCAI (2018)

\bibitem{elsayed2024hmar}
Elsayed, S., Rashed, A., Schmidt-Thieme, L.: Hmar: Hierarchical masked attention for multi-behaviour recommendation. In: PAKDD (2024)

\bibitem{gao2019neural}
Gao, C., He, X., Gan, D.: Neural multi-task recommendation from multi-behavior data. In: ICDE (2019)

\bibitem{gu2022self}
Gu, S., Wang, X., Chi, C., Xiao, D.: Self-supervised graph neural networks for multi-behavior recommendation. In: IJCAI (2022)

\bibitem{guo2019buying}
Guo, L., Hua, L., Jia, R., Zhao, B., Wang, X., Cui, B.: Buying or browsing?: Predicting real-time purchasing intent using attention-based deep network with multiple behavior. In: KDD (2019)

\bibitem{han2024efficient}
Han, Y., Wang, H., Wang, K., Wu, L., Li, Z., Guo, W., Liu, Y., Lian, D., Chen, E.: Efficient noise-decoupling for multi-behavior sequential recommendation. In: WWW (2024)

\bibitem{he2020lightgcn}
He, X., Deng, K., Wang, X., Li, Y., Zhang, Y., Wang, M.: Lightgcn: Simplifying and powering graph convolution network for recommendation. In: SIGIR (2020)

\bibitem{hu2008collaborative}
Hu, Y., Koren, Y., Volinsky, C.: Collaborative filtering for implicit feedback datasets. In: ICDM (2008)

\bibitem{jin2020multi}
Jin, B., Gao, C., He, X., Jin, D., Li, Y.: Multi-behavior recommendation with graph convolutional networks. In: SIGIR (2020)

\bibitem{kipf2016semi}
Kipf, T.N., Welling, M.: Semi-supervised classification with graph convolutional networks. In: ICLR (2017)

\bibitem{koren2008factorization}
Koren, Y.: Factorization meets the neighborhood: a multifaceted collaborative filtering model. In: KDD (2008)

\bibitem{Krohn-Grimberghe2012multi}
Kronh-Grimberghe, A., Drumond, L., Freudenthaler, C., Schmidt-Thieme, L.: Multi-relational matrix factorization using bayesian personalized ranking for social network data. In: WSDM (2012)

\bibitem{lee2024mule}
Lee, S., Ko, G., Song, H.J., Jung, J.: Mule: Multi-grained graph learning for multi-behavior recommendation. In: CIKM (2024)

\bibitem{li2023graph}
Li, C., Xia, L., Ren, X., Ye, Y., Xu, Y., Huang, C.: Graph transformer for recommendation. In: SIGIR (2023)

\bibitem{li2018learning}
Li, Z., Zhao, H., Liu, Q., Huang, Z., Mei, T., Chen, E.: Learning from history and present: Next-item recommendation via discriminatively exploiting user behaviors. In: KDD (2018)

\bibitem{lin2022improving}
Lin, Z., Tian, C., Hou, Y., Zhao, W.X.: Improving graph collaborative filtering with neighborhood-enriched contrastive learning. In: WWW (2022)

\bibitem{liu2024multi}
Liu, Z., Hou, Y., McAuley, J.: Multi-behavior generative recommendation. In: CIKM (2024)

\bibitem{loni2016bayesian}
Loni, B., Pagano, R., Larson, M., Hanjalic, A.: Bayesian personalized ranking with multi-channel user feedback. In: RecSys (2016)

\bibitem{luo2022dual}
Luo, J., He, M., Lin, X., Pan, W., Ming, Z.: Dual-task learning for multi-behavior sequential recommendation. In: CIKM (2022)

\bibitem{luo2023criterion}
Luo, X., Wu, D., Gu, Y., Chen, C., Liu, L., Ma, J., Zhang, M., Deng, M., Huang, J., Hua, X.S.: Criterion-based heterogeneous collaborative filtering for multi-behavior implicit recommendation. ACM Transactions on Knowledge Discovery from Data  \textbf{18}(14),  1--26 (2023)

\bibitem{ma2022crosscbr}
Ma, Y., He, Y., Zhang, A., Wang, X., Chua, T.S.: Crosscbr:cross-view contrastive learning for bundle recommendation. In: KDD (2022)

\bibitem{meng2023coarse}
Meng, C., Zhao, Z., Guo, W., Zhang, Y., Wu, H., Gao, C., Li, D., Li, X., Tang, R.: Coarse-to-fine knowledge-enhanced multi-interest learning framework for multi-behavior recommendation. ACM Transactions on Information Systems  \textbf{42}(30),  1--27 (2023)

\bibitem{meng2023parallel}
Meng, C., Zhai, C., Yang, Y., Zhang, H., Li, X.: Parallel knowledge enhancement based framework for multi-behavior recommendation. In: CIKM (2023)

\bibitem{meng2023hierarchical}
Meng, C., Zhang, H., Guo, W., Guo, H., Liu, H., Zhang, Y., Zheng, H., Tang, R., Li, X., Zhang, R.: Hierarchical projection enhanced multi-behavior recommendation. In: KDD (2023)

\bibitem{qiu2018bprh}
Qiu, H., Liu, Y., Guo, G., Sun, Z., Zhang, J., Nguyen, H.T.: Bprh: Bayseian personalized ranking for heterogeneous implicit feedback. Information Sciences  \textbf{453},  80--98 (2018)

\bibitem{singh2008relational}
Singh, A.P., Gordon, G.J.: Relational learning via collective matrix factorization. In: KDD (2008)

\bibitem{su2023personalized}
Su, J., Chen, C., Lin, Z., Li, X., Liu, W., Zheng, X.: Personalized behavior-aware transformer for multi-behavior sequential recommendation. In: MM (2023)

\bibitem{sun2019bert4rec}
Sun, F., Liu, J., Wu, J., Pei, C., Lin, X., Ou, W., Jiang, P.: Bert4rec: Sequential recommendation with bidirectional encoder representations from transformer. In: CIKM (2019)

\bibitem{wei2022constrastive}
Wei, W., Huang, C., Xia, L., Xu, Y., Zhao, J., Yin, D.: Contrastive meta learning with behavior multiplicity for recommendation. In: WSDM (2022)

\bibitem{wei2021contrastive}
Wei, Y., Wang, X., Li, Q., Nie, L., Li, Y., Li, X., Chua, T.S.: Contrastive learning for cold-start recommendation. In: MM (2021)

\bibitem{Wu2022DCRec}
Wu, J., Fan, W., Chen, J., Liu, S., Li, Q., Tang, K.: Disentangled contrastive learning for social recommendation. In: CIKM (2022)

\bibitem{xia2023automated}
Xia, L., Huang, C., Huang, C., Kangyi, L., Yu, T., Ben, K.: Automated self-supervised learning for recommendation. In: WWW (2023)

\bibitem{xia2021multi}
Xia, L., Huang, C., Xu, Y., Dai, P., Lu, M., Bo, L.: Multi-behavior enhanced recommendation with cross-interaction collaborative relation modeling. In: ICDE (2021)

\bibitem{xia2020multiplex}
Xia, L., Huang, C., Xu, Y., Dai, P., Zhang, B., Bo, L.: Multiplex behavioral relation learning for recommendation via memory augmented transformer network. In: SIGIR (2020)

\bibitem{xia2021knowledge}
Xia, L., Huang, C., Xu, Y., Dai, P., Zhang, X., Yang, H., Pei, J., Bo, L.: Knowledge-enhanced hierarchical graph transformer network for multi-behavior recommendation. In: AAAI (2021)

\bibitem{xia2021graph}
Xia, L., Xu, Y., Huang, C., Dai, P., Bo, L.: Graph meta network for multi-behavior recommendation. In: SIGIR (2021)

\bibitem{xin2023improving}
Xin, X., Liu, X., Wang, H., Ren, P., Chen, Z., Lei, J., Shi, X., Luo, H., Jose, J.M., de~Rijke, M., Ren, Z.: Improving implicit feedback-based recommendation through multi-behavior alignment. In: SIGIR (2023)

\bibitem{xu2023multi}
Xu, J., Wang, C., Wu, C., Song, Y., Zheng, K., Wang, X., Changping, W., Zhou, G., Gai, K.: Multi-behavior self-supervised learning for recommendation. In: SIGIR (2023)

\bibitem{xuan2023knowledge}
Xuan, H., Liu, Y., Li, B., Yin, H.: Knowledge enhancement for contrastive multi-behavior recommendation. In: WSDM (2023)

\bibitem{yan2023cascading}
Yan, M., Chen, Z., Gao, C., Sun, Z., Liu, F., Sun, F., Li, H.: Cascading residual graph convolutional network for multi-behavior recommendation. ACM Transactions on Information Systems  \textbf{1},  1--24 (2023)

\bibitem{yan2023mbhgcn}
Yan, M., Cheng, Z., Sun, J., Sun, F., Peng, Y.: Mb-hgcn: A hierarchical graph convolutional network for multi-behavior recommendation. arXiv preprint arXiv:2306.10679  (2023)

\bibitem{yan2024behavior}
Yan, M., Liu, F., Sun, J., Sun, F., Chen, Z., Han, Y.: Behavior-contextualized item preference modeling for multi-behavior recommendation. In: SIGIR (2024)

\bibitem{yang2021hyper}
Yang, H., Chen, H., Li, L., Yu, P.S., Xu, G.: Hyper meta-path contrastive learning for multi-behavior recommendation. In: ICDM (2021)

\bibitem{yang2022knowledge}
Yang, Y., Huang, C., Xia, L., Li, C.: Knowledge graph contrastive learning for recommendation. In: SIGIR (2022)

\bibitem{yang2022multi}
Yang, Y., Huang, C., Xia, L., Liang, Y., Yu, Y., Li, C.: Multi-behavior hypergraph-enhanced transformer for sequential recommendation. In: KDD (2022)

\bibitem{wu2022multi}
Yiqing, W., Ruobing, X., Yongchun, Z., Xiang, A., Xin, C., Xu, Z., Fuzhen, Z., Leyu, L., Qing, H.: Multi-view multi-behavior contrastive learning in recommendation. In: DASFAA (2022)

\bibitem{yu2022graph}
Yu, B., Zhang, R., Chen, W., Fang, J.: Graph neural network based model for multi-behavior session-based recommendation. GeoInformatica  \textbf{26}(2),  429--447 (2022)

\bibitem{yu2023xsimgcl}
Yu, J., Xia, X., Chen, T., Cui, L., Hung, N.Q.V., Yin, H.: Xsimgcl: Towards extremely simple graph contrastive learning for recommendation. IEEE Transactions on Knowledge and Data Engineering  \textbf{36}(2),  913--926 (2023)

\bibitem{yuan2022multi}
Yuan, E., Guo, W., He, Z., Guo, H., Liu, C., Tang, R.: Multi-behavior sequential transformer recommender. In: SIGIR (2022)

\bibitem{zhai2024actions}
Zhai, J., Liao, L., Liu, X., Wang, Y., Li, R., Cao, X., Gao, L., Gong, Z., Gu, F., He, M., Lu, Y., Shi, Y.: Actions speak louder than words: Trillion-parameter sequential transducers for generative recommendations. In: ICML (2024)

\bibitem{zhang2019heterogeneous}
Zhang, C., Song, D., Huang, C., Swami, A., Chawla, N.V.: Heterogeneous graph neural network. In: KDD (2019)

\bibitem{zhang2020multiplex}
Zhang, W., Mao, J., Cao, Y., Xu, C.: Multiplex graph neural networks for multi-behavior recommendation. In: CIKM (2020)

\bibitem{zhang2024multi}
Zhang, Y., Bei, Y., Chen, H., Shen, Q., Yuan, Z., Gong, H., Wang, S., Huang, F., Huang, X.: Multi-behavior collaborative filtering with partial order graph convolutional networks. In: KDD (2024)

\bibitem{zhou2019collaborative}
Zhou, X., Liu, D., Lian, J., Xie, X.: Collaborative metric learning with memory network for multi-relational recommender systems. In: IJCAI (2019)

\bibitem{zhu2024multi}
Zhu, X., Lin, F., Zhao, Z., Xu, T., Zhao, X., Yin, Z., Li, X., Chen, E.: Multi-behavior recommendation with personalized directed acyclic behavior graphs. ACM Transactions on Information Systems  \textbf{43}(1),  1--30 (2024)

\end{thebibliography}

\end{document}